\documentstyle[prl,aps]{revtex}

\newcommand{\F}{{\cal F}}

\newcommand{\h}{{\cal H}}
\newcommand{\Q}{{\cal Q}}

\newcommand{\be}{\begin{equation}}
\newcommand{\ee}{\end{equation}}
\newcommand{\ben}{\begin{enumerate}}
\newcommand{\een}{\end{enumerate}}

\begin{document}

\input{epsf}
\draft

\twocolumn[\hsize\textwidth\columnwidth\hsize\csname @twocolumnfalse\endcsname
  
\title{Real-Space Renormalization Group Study of the
Anisotropic Antiferromagnetic Heisenberg Model
on the Square Lattice \thanks{CAPES}}

\author{ \sc N. S. Branco \thanks{Permanent address:
Depto.\ de F\'{\i}sica, Universidade Federal de Santa
Catarina, 
88040-900, Florian\'opolis, SC  - Brazil; e-mail:
nsbranco@fisica.ufsc.br} }
\address{Center for Simulational Physics \\ Department of Physics
and Astronomy - University of Georgia\\ 
Athens, GA 30602, USA\\ e-mail:
nsbranco@hal.physast.uga.edu }
\author{ J.\ Ricardo de Sousa }
\address{ Departamento de F\'{\i}sica - Universidade do Amazonas\\
3000 Japiim,
69077-000, Manaus, AM  - Brazil\\   e-mail: jrs@fisica.fua.br }

\date{\today}

\maketitle

\begin{abstract}

 	In this work we apply two different real-space
renormalization-group (RSRG) approaches to the anisotropic
antiferromagnetic spin-1/2
Heisenberg model on the square lattice. Our calculations
allow for an approximate evaluation of the $T$ vs. $\Delta$
phase-diagram: the results suggest the existence of a 
critical value of 
$\Delta>0$, at which the critical temperature goes to zero,
and  the presence of reentrant behavior on the critical line between 
the ordered and disordered phases. This
whole critical line is found to belong to the same universality
class as the Ising model. Our results are in accordance with previous
RSRG approaches but not with numerical simulations
and spin-wave calculations.

\end{abstract}

\pacs{64.60.Ak; 75.50.Ee; 74.72.Dn}

\vskip2pc]   

\section{Introduction}
            
	 Two-dimensional antiferromagnetism plays an important
role in the description of La$_2$CuO$_4$-based high-temperature
superconductors \cite{shirane}. In these, the spin fluctuations
in the CuO$_2$ planes are well described by the spin-1/2 
antiferromagnetic Heisenberg model on a square lattice.
This, together with the theoretical discussion
made by Anderson \cite{anderson},
has raised the interest on this model, which has been a challenge
for decades.

	The Hamiltonian of the anisotropic Heisenberg model reads:
\be
   - \beta \h = K \sum_{\left<i,j\right>}
      \left[ (1-\Delta) \left( S^x_i S^x_j + S^y_i S^y_j \right)
      + S^z_i S^z_j \right],   \label{hamil}
\ee
where $K=J/kT$ is the dimensionless exchange parameter, 
with $K<0(>0)$ for the 
antiferromagnetic (ferromagnetic) model, $k$ is the Boltzmann
constant, $T$ is the temperature, the sum is over
pairs of nearest-neighbor spins $i$ and $j$, $S_l^{\lambda}$
is the $\lambda^{\mbox{th}}$ component of the spin-1/2 Pauli operator
on site $l$, $\Delta$ is the anisotropy parameter, and $\beta = 1/kT$.
Note that $\Delta=0$ describes the isotropic
Heisenberg model and for $\Delta=1$ we regain the Ising model.

\begin{figure}
\epsfxsize=6.5cm
\begin{center}
\leavevmode
\epsffile{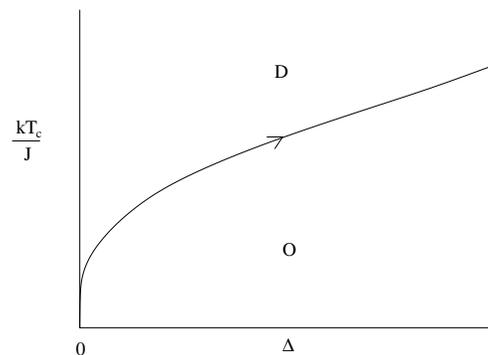}
\caption{Schematic plot of the temperature vs. anisotropy 
phase-diagram for the two-dimensional anisotropic ferromagnetic 
Heisenberg model,
where $O$ stands for the ferromagnetic (ordered) phase and $D$ stands
for the paramagnetic (disordered) phase.
The arrow indicates that the transitions for $\Delta \neq 0$
belong to the same universality class as for the Ising model
($\Delta=1$). Note that the critical temperature line goes to zero
in the limit of the isotropic Heisenberg model ($\Delta=0$).}
\label{ferro} 
\end{center}
\end{figure}

	For $J>0$ (ferromagnetic interactions), the ground state
is well known and the Mermin-Wagner theorem excludes long-range
order at finite temperatures in two dimensions for
$\Delta=0$ \cite{mermin}. On the other hand, for 
$0 < \Delta \leq 1$ there is an
easy-axis and the symmetry is the same as for the Ising model;
therefore, long-range order is possible. Indeed, the universality
class for all models with values of $\Delta \neq 0$ is the same
as for the Ising model. A schematic temperature vs. anisotropy 
phase-diagram
for the ferromagnetic model is presented in Fig.\ref{ferro},
where all the features discussed above are exhibited.

	For many classical (and some quantum) systems on bipartite 
lattices, there is a mapping of the ferromagnetic model onto the  
antiferromagnetic one. Nevertheless, there is 
no such mapping for the Heisenberg model; therefore, the 
ground state of the anisotropic antiferromagnetic Heisenberg
(AAH) model is not obtained from its ferromagnetic 
couterpart by flipping all spins in a given sub-lattice. In fact, the
ground state for $J<0$ is not even exactly known and has been a
matter of debate for a long time. However, long-range order was
proven to be present
at $T=0$ for $\Delta > \Delta_c = 0.40$ \cite{nishimori}, whereas 
the same authors, using extra assumptions, calculated a lower value 
of $\Delta_c$, namely, $\Delta_c = 0.09$ \cite{nishimori}.

	Previous RSRG procedures have been able to calculate
the approximate phase-diagram for the AAH on the square lattice
\cite{andre,ricardo}. The former reference uses a hierarchical
lattice to approximate the square one, and performs a partial
trace over internal degrees of freedom, in a manner introduced 
in Ref. \onlinecite{ananias}. This approach was the extension of the
Niemeijer-van Leeuwen method to quantum spin systems. On the other
hand, Ref. \onlinecite{ricardo} applies the so-called mean-field 
renormalization group
ideas. Although the approximate scaling transformations calculated 
in  Refs.\cite{andre} and \cite{ricardo} are different, the results 
obtained are qualitatively the same, and we will comment on them 
when our results are discussed. Nevertheless, one should bear in
mind that RSRG procedures on 
quantum systems do not have the same firm basis as on classical
models. The reason is the non-commutivity aspects of the
Hamiltonian and, in as concerns Ref. \onlinecite{ricardo}, the
necessity of introducing symmetry-breaking fields which are chosen
according to the ground state of the {\it classical} Ising model.

	Results from spin-wave theory \cite{spin} on the
isotropic Heisenberg model give the same value for the critical 
temperature of the
ferromagnetic {\it and} antiferromagnetic systems \cite{ricardo2}, 
which is in direct contradiction to rigorous results \cite{rush}
and to some approximate calculations \cite{andre,ricardo}.

	Therefore, the form of the phase-diagram of the AAH model
on the square lattice is far from a settled question and more work
is desirable to study its critical properties. We thus apply 
two different RSRG procedures to evaluate the approximate 
phase-diagram of this system.
The ferromagnetic Heisenberg case was also studied, in order 
to compare our results
with previous ones for a model for which the critical behavior is
well known.

	The remainder of this paper is organized as follows. 
In section II we outline the formalism we used, 
in section III we present results, and in the last 
section we summarize our main conclusions.  

\section{Formalism}

	In this section we outline the finite-size scaling RG (FSSRG)
procedure, since it is somewhat new in the literature 
and, to the best of our knowledge,
has never been used in the study of quantum systems. 
This method was proposed some time ago \cite{murilo} and has been
successfully applied to classical systems (both static and dynamic
properties have been studied), like Ising, Potts, or Blume-Capel
models \cite{pla}. The second is just the generalization of the usual
bond-moving Migdal-Kadanoff \cite{migkad}
approximation to antiferromagnetic quantum systems.

	The finite-size scaling assumption is that, near the critical
region, thermodynamic quantities have the following form
\cite{barber}:
\be
     \F \left( \varepsilon, L \right)  = b^{\psi} 
     \F \left( b^{1/\nu} \varepsilon, b^{-1} L \right),
\ee
where $b$ is some arbitrary scaling factor,
$L$ is the linear dimension of the lattice, 
$\F$ is a scaling function,
$\varepsilon \equiv \left| T-T_c \right|$ ($T_c$ is the critical
temperature), and $\nu$ is the critical exponent of the
correlation length, such that $\xi \sim \varepsilon^{-\nu}$
for $T$ close to $T_c$ and in the thermodynamic limit. The exponent
$\psi$ is the anomalous dimension of the thermodynamic quantity
$\F$; for
the magnetization $M$, $\psi=-\beta/\nu$, while for the
magnetic susceptibility $\chi$, $\psi=\gamma/\nu$.

	The idea behind the FSSRG is to construct quantities which
have a zero anomalous dimension, $\psi=0$, such that:
\be  
   \Q_{L'}(\varepsilon') \equiv \Q_{L'}(b^{1/\nu} \varepsilon) 
   = \Q_L (\varepsilon). \label{fss}
\ee
As seen, these quantities will have the same value at $T=T_c$,
no matter what the lattice sizes $L$ or $L'$ are (as far as 
both are $\gg 1$). Therefore,
the crossing of $\Q$ for two different lattice sizes is an evaluation
of the critical temperature; furthemore, the previous equation
can be seen as an iteration, in the renormalization group (RG)
sense. Thus, information
on the exponent $\nu$ can also be accessed, as well as
other information obtained from RG procedures (universality
classes, crossover phenomena, first order phase transitions,
etc). For example, for the Ising model in zero magnetic field,
one apropriate function is:
\be
     \tau = \left< \mbox{sign} \left( \sum_{top} S_t \right)
   \mbox{sign} \left( \sum_{bottom} S_b \right) \right>
\ee
where $\left< \ldots \right>$ denotes a canonical average, $bottom$ 
means
all spins at the bottom plane(line) of a three(two)-dimensional
lattice, $top$ stands for all spins at the top plane(line) of
a three(two)-dimensional lattice, and $\mbox{sign}(x)=-1, 0$
or $1$ if $x<0$, $x=0$ or $x>0$, respectively. For an infinite
system, $\tau=1$ for $T<T_c$ and $\tau=0$ for $T>T_c$. Replacing 
$\tau$ for $\Q$ in Eq. \ref{fss}, we obtain a RSRG equation which
connects the parameters $K$ in the original lattice and $K'$ in the
renormalized (smaller) 
lattice. For a more detailed discussion on the application
of the FSSRG to the Ising model, see Ref. \onlinecite{murilo}.
Note, however, that the only approximation
comes from the finite size of the lattices; in fact, even
when small lattices are used, the quantitative results are
precise, and this can be understood if one realizes that the FSSRG
is a generalization, for completely finite clusters,
of the phenomenological RG developed by Nightingale \cite{night}. 

\begin{figure}
\epsfxsize=6.5cm
\begin{center}
\leavevmode
\epsffile{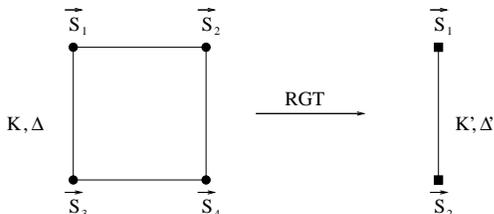}
\caption{Cells used to implement the renormalization group
transformation for the anisotropic Heisenberg model on
the square lattice.}
\label{cells} 
\end{center}
\end{figure}

	In using the FSSRG approach in the study of quantum systems,
we expect to take advantage of the fact that non-commutative aspects
of the Hamiltonian are not approximated away by the method.
Our approach here is to use small lattices, which, although not
allowing for a precise evaluation of the critical parameters
(like, for instance, those obtained with numerical simulations
on big lattices), allows for a good qualitative description of the
critical phenomena involved. The lattices we chose to represent
the square lattice are depicted in Fig. \ref{cells}, where the right 
figure is the bigger cluster,
with parameters $K \equiv J/kT$ and $\Delta$ and the left figure
depicts the smaller cluster, with parameters $K'$ and
$\Delta'$. The Hamiltonian for the original and renormalized clusters
are:
\begin{eqnarray}
   -\beta \h & = & K \left[ (1-\Delta) ( S^x_1 S^x_2 + 
       S^y_1 S^y_2 + S^x_2 S^x_3 + S^y_2 S^y_3 + \right.
         \cr\cr
   &  &  S^x_3 S^x_4 + S^y_3 S^y_4 +
    + S^x_4 S^x_1 + S^y_4 S^y_1 ) +  \cr\cr 
  &  & 
    \left. S^z_1 S^z_2 + S^z_2 S^z_3 + S^z_3 S^z_4 + S^z_4 S^z_1 \right], 
   \label{oldhamil} 
\end{eqnarray} 
and
\be
    -\left( \beta \h \right)' = K' \left[ (1-\Delta') \left( 
      S^x_1 S^x_2 + S^y_1 S^y_2 \right) + S^z_1 S^z_2 \right], 
      \label{newhamil}
\ee
respectively.

	When applying the FSSRG method to the anisotropic 
Heisenberg model, one has to devise two functions with zero 
anomalous dimension, since it is necessary to renormalize two
parameters, namely, $\Delta$ and $kT/J$. Also, we have to
distinguish between the antiferromagnetic and ferromagnetic models,
and two different sets of functions were used for each of them.

	For the ferromagnetic model, we chose the following
quantities:
\be
    \tau' = \left< \mbox{sign} (S_1^z) \mbox{sign} (S_2^z)
        \right> ;   \label{tau'}
\ee
\be
    \eta' = \left< \mbox{sign} (S_1^x) \mbox{sign} (S_2^x)
        \right>      \label{eta'}
\ee
for the smaller cluster and
\be
    \tau =  \left< \mbox{sign} (S_1^z+S_2^z) \mbox{sign}
        (S_3^z+S_4^z) \right>  \label{tauf}
\ee
\be
     \eta =  \left< \mbox{sign} (S_1^x+S_2^x) \mbox{sign}
        (S_3^x+S_4^x) \right>   \label{etaf}
\ee
for the bigger cluster. The averages are to be taken with respect
to the ensemble defined by Eq. \ref{oldhamil} (\ref{newhamil})
for unprimed (primed) quantities.

	For the antiferromagnetic model, Eqs. \ref{tau'} and
\ref{eta'} remain the same, while Eqs. \ref{tauf} and \ref{etaf}
are replaced by:
\be
    \tau =  \left< \mbox{sign} (S_1^z-S_2^z) \mbox{sign}
        (-S_3^z+S_4^z) \right>  \label{taua}
\ee
and
\be
   \eta =  \left< \mbox{sign} (S_1^x-S_2^x) \mbox{sign}
        (-S_3^x+S_4^x) \right>,  \label{etaa}
\ee
respectively.

	The procedure to calculate these functions is fairly
straightforward; the final expressions, however, are too lengthy
and will be omitted. To obtain the required RG equations, we 
impose $\tau'=\tau$ and $\eta'=\eta$, according to Eq. \ref{fss}.
Approximate values for the critical points are obtained from the 
fixed points of these equations and critical exponents are
linked to the behavior of the iterations near the fixed points.

\section{Results}

	\subsection{Finite-size scaling renormalization 
group (FSSRG)}

\begin{figure}
\epsfxsize=6.5cm
\begin{center}
\leavevmode
\epsffile{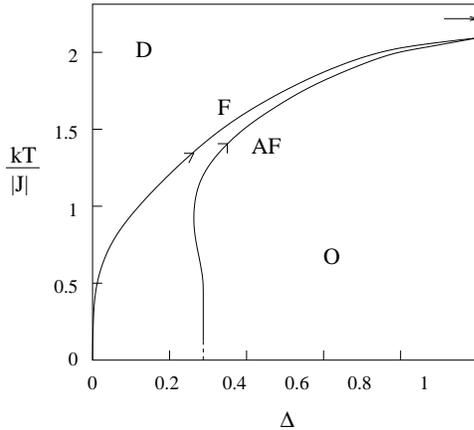}
\caption{Critical temperature vs. anisotropy
phase-diagram for the ferromagnetic (curve $F$) and 
anti-ferromagnetic (curve $AF$) anisotropic Heisenberg model. Both 
lines are atracted to the Ising fixed-point, located at
$kT/|J|=2.11, \Delta=1$. $O$($D$) stands for ordered (disordered)
phase. The arrow indicates the exact critical temperature for
the Ising model.}
\label{FSS} 
\end{center}
\end{figure}

	Our results for the ferromagnetic (F) and antiferromagnetic 
(AF) models are shown in Fig. \ref{FSS}. The critical curve for the 
F curve is depicted for comparison: the accordance with previous
results, either approximate or exact, is very good. The Ising critical
point is located at $kT_c/|J|=2.11, \Delta=1$ (exact values:
$kT_c/|J|=2.269, \Delta=1$); our estimate for the critical exponent
$\nu$ for the Ising model is 0.91, while the exact value is 1. 
The universality
class for $\Delta \neq 0$ is the same as for the Ising model, which
is also consistent with previous results. 
Note that the F and AF Ising models are expected to have the same 
critical
exponents and the same modulus of the critical temperature, and our
evaluation agrees with these results. 

	For the AF model, the phase diagram is qualitatively different
from its F counterpart: the critical temperature reaches zero at a
critical value of $\Delta$, $\Delta_c$, which is greater than
zero. We find $\Delta_c=0.29$, which compares with
$\Delta_c=0.40$ in Ref. \onlinecite{andre} and $\Delta_c=0.18$ in
Ref. \onlinecite{ricardo}. We also find a reentrant behavior in the
critical line, which is also present in Refs. \cite{andre,ricardo}.
Nevertheless, in Ref. \onlinecite{ricardo} a second reentrance is
observed; the lowest temperature we could work with was $kT/|J|=0.1$
and we have observed no sign of this second reentrance, neither with
the FSSRG nor with the bond-moving scheme (to be presented in what 
follows). In Ref. \onlinecite{ricardo}, the N\'eel temperature
behaves as $T_N \sim 1/\log(\Delta-\Delta_c)$
near $\Delta=\Delta_c$, while, for the anisotropic
ferromagnetic Heisenberg model, $T_c \sim 1/\log(\Delta)$ 
\cite{ricardo}.
The supression of long-range order at finite temperatures for
small values of $\Delta$ can be regarded as due to quantum 
fluctuations. While these are not relevant in critical phenomena 
which take place at ``high'' temperatures, they might be
important when the critical temperature is low. We expect 
this to be the case for small values of $\Delta$, and then quantum
fluctuations gain in importance, supressing long-range order.
Note that we cannot present results at $T=0$, since parts of our
calculation were done numerically and, therefore, we are not
able to go to very low temperatures.

	\subsection{Migdal-Kadanoff approximation}

	We have also performed a bond-moving RSRG procedure to
study the AAH model on the square lattice. This procedure is
equivalent to the one in Ref. \onlinecite{andre} applied to a 
different hierarchical lattice; a careful exposition of the 
method is made in Ref. \onlinecite{ananias} and the bond-moving 
approximation is presented in Ref. \onlinecite{anna}.

\begin{figure}
\epsfxsize=6.5cm
\begin{center}
\leavevmode
\epsffile{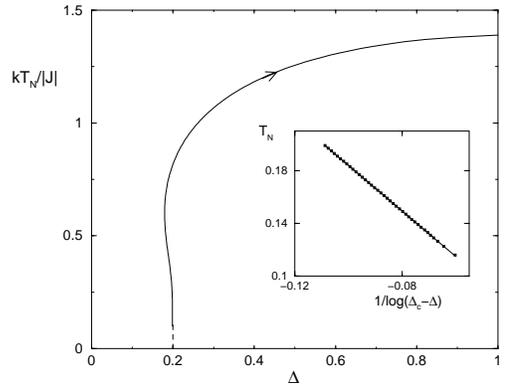}
\caption{N\'eel temperature $T_N$ vs. $\Delta$ phase-diagram 
for the antiferromagnetic anisotropic Heisenberg model on the square
lattice. The region
above (below) the critical line represents a(n) desordered
(ordered) phase and the insert shows the data for small $T_N$,
near $\Delta_c=0.199$.}
\label{mk} 
\end{center}
\end{figure}

	The phase-diagram for the antiferromagnetic model is presented in 
Fig. \ref{mk}; note that the qualitative features agree with those
obtained from other RSRG approaches. However, we have not found the
second reentrance in the critical curve. The N\'eel temperature $T_N$ 
varies as (see insert in Fig. \ref{mk}):
\be
    T_N \sim \frac{1}{\log \left( \Delta_c-\Delta \right)},
\ee
where $\Delta_c=0.199$ for the Migdal-Kadanoff approximation. 
Again the universality class for the whole critical curve is 
the same as for the Ising model.

	We would like to mention that the results from both
RSRG employed here are in disagreement
with spin-wave calculations \cite{spin,ricardo2} and numerical 
simulation \cite{russo,MC}. The former presents $T_c=T_N$ for any 
value of $\Delta$ and for two and three dimensions, while the latter
predicts $\Delta_c=0$ for the AAH model. In Ref. \onlinecite{MC}, the
logarithmic dependence of $T_N$ and $T_c$ with respect to
$\Delta-\Delta_c$ is established using scaling arguments, but
$\Delta_c=0$ in that paper. Note that our calculations
cannot be carried out down to $T=0$; therefore, we cannot
study the character of the ground state of the AAH model
\cite{nishimori,heisenberg}.

\section{Summary}

	In summary, we calculate the $kT/|J|$ vs. $\Delta$ 
phase-diagram
for the anisotropic antiferromagnetic Heisenberg model on the square
lattice. Our results show the presence of reentrant behavior on the
critical line which separates the disordered and ordered phases
and a value of $\Delta>0$ such that the N\'eel temperature is zero.
The entire critical line is found to belong to the universality class
of the Ising model. These findings are in agreement with previous
RSRG procedures (with the exception of the second reentrance found
in Ref. \onlinecite{ricardo}) but not with spin-wave calculations and
numerical simulation.

	It is clear that the reentrant behavior and a value of 
$\Delta_c$
greater than zero are strongly supported by RSRG approaches, but the 
question is not yet settled and more work is needed to put these
points on firmer grounds. Coherent-anomaly methods \cite{suzuki} and 
numerical simulations  with cluster algorithms \cite{algoritmo} are 
possible ways to  provide a more definite answer to this problem.
Work is now proceeding along these lines.

\section{Acknowledgments}

   	We would like to thank Dr. J. A. Plascak and Prof. D. P.
Landau for discussions and for a critical reading of the manuscript.

\end{document}